\begin{filecontents*}{}
\documentclass [a4paper,11pt]{article}
\usepackage[english]{babel}
\usepackage{graphicx,amsmath,amsfonts,amsbsy, amssymb,slashed,upgreek,color,caption,amscd,cite,cancel,enumerate,relsize}

\def\bs{\boldsymbol}

\def\d{\operatorname{d}\!}
\def\pd{\partial}

\def\hL{\mathcal{H}_{\Lambda}}

\begin{document}

\begin{center}
\Large{\textbf{On the dynamical emergence of de Sitter spacetime}}\\[1cm]

\large{Christian Marinoni$^{\rm a,b}$ and Heinrich Steigerwald$^{\rm a}$}
\\[0.5cm]

\small{
\textit{$^{\rm a}$  Aix-Marseille Universit\'e, Universit\'e de Toulon, CNRS, CPT UMR 7332, 13288, Marseille, France
}}\\[1mm]

\small{
\textit{$^{\rm b}$ Institut Universitaire de France, 103, bd. Saint-Michel, F-75005 Paris, France}}

\vspace{.2cm}

\end{center}

\vspace{2cm}

\begin{abstract}
We present and discuss an asynchronous  coordinate system covering de Sitter spacetime, notably in a complete way  in 1+1 dimensions. The new coordinates  have several   interesting 
cosmological properties: the worldlines of comoving ($x^i=const$)  observers are geodesics, 
cosmic time is finite in the past,  and the coordinates asymptotically tend   to that of a flat  Robertson \& Walker model  at large times.
This analysis also provides an argument in favor of the natural emergence  of an equation of state of the type  $p=-\rho$  in the context of  the standard cosmological model. 
\end{abstract}

\vspace{9cm}

\begin{footnotesize}
E-mail:  \\
\tt{christian.marinoni@cpt.univ-mrs.fr\\
heinrich.steigerwald@cpt.univ-mrs.fr \\ 
}
\end{footnotesize}

\newpage
\tableofcontents
\vspace{.5cm}
\newpage

\section{Introduction}

De Sitter spacetime   \cite{des1,des2}  plays a central role in cosmology.  Its special status is justified on  both historical and physical  grounds. 
It was the  first  expanding world model ever proposed \cite{lan}, the very one predicting the cosmological redshift of light, and the very one, together with Minkowski and anti-de Sitter spacetimes, to embody 
the maximal degree of symmetry  in both space and time.  Despite its idealistic nature --
this spacetime cannot accommodate matter --  the de Sitter model is still nowadays an essential framework to  understand the critical properties of the primordial inflationary universe 
as well as the asymptotic  future state of a universe dominated by the cosmological constant.
Indeed,  during both these epochs, the evolution of the universe can be effectively modelled as a de Sitter stage with slightly broken time-translational symmetries.

There are many equivalent geometrical  definitions of  a de Sitter spacetime, each of them illustrating and emphasising complementary  facets of its metric.
It can  be defined, as the spacetime admitting  the de Sitter group $O(4,1)$ as group of symmetries \cite{schro}, {\it i.e.},  the 4-dimensional spacetime
that results from  embedding  a hyperboloid in a 5-dimensional Minkowski spacetime.  It can also be obtained via the analytical continuation of a metric describing 
a positive curvature space with Euclidean signature into a constant curvature space of Lorentzian signature \cite{muk}.
In a cosmological context,  the dynamical emergence of the de Sitter spacetime as  the vacuum solution to the Einstein's Field Equations (EFE) with a positive cosmological constant \cite{des1,des2} is of more interest. In other terms, the existence of  a cosmological component  with an effective energy density $\rho$ and pressure $p$ related by an equation of state $p=-\rho$ is mandatory if we are to obtain the de Sitter spacetime as solution of the EFE. 

A large variety of coordinate systems is  known for de Sitter space (see, e.g., \cite{eg,kyp,mosca}). 
In the context of General Relativity,  because of local diffeomorphism invariance,  coordinate systems play no role in the formulation of fundamental physical laws, only diffeomorphism invariants matter. On the other hand, a poor choice of local coordinates can sometimes obscure the interpretation of physical phenomena and the fundamental properties of spacetime.  As an example, suitably  tailored coordinate systems, which embody specific symmetries of a physical system, may  help to solve  differential equations \cite{mosca2}.  Furthermore, coordinate systems that cover larger patches of spacetime are specifically  useful when dealing with physical phenomena that are  non-localized \cite{Parikh} or  which are sensitive to  global properties of spacetime
 \cite{GMNV}. 

The purpose of this paper is to discuss the cosmological properties of a specific  comoving geodesic coordinate system, describing a 3D space with  translational and rotational symmetries,
which admits the flat Robertson \& Walker (R\&W) coordinate system as  future asymptotic attractor.  Moreover the only cosmological solution of the EFE  (without the cosmological term) 
corresponds to a  perfect fluid component  with an effective equation of state  $p=-\rho$. In other terms,  more than  a mere  option in the space of the possible outcomes of cosmological experiments,  such an  `exotic'  equation of state naturally emerges as a necessary  ingredient of  the universe for  a specific class of comoving geodesic observers which becomes R\&W at later epochs.  

Throughout this paper we suppose that cosmic spacetime  is a  Rienmanian manifold $(\mathcal{M},g)$ and we adopt the sign convention $(+,-,-,-)$ for the metric $g$. If not indicated otherwise, we adopt Einstein's convention summing over repeated  indices. Greek indices run over spacetime coordinates (from 0 to 3) and latin indices run over space coordinates (from 1 to 3). In particular, $x^i=(x,y,z)$ indicate the standard cartesian coordinates. We set the light speed $c=1$ but we keep the gravitational constant $G$ explicit in the equations.

%
\section{A new class of comoving observers}
In this section we introduce  the cosmological coordinate system $x^{\mu}$ that is central to our analysis. Specifically, we show that one can define a class of  comoving observers ($x^i=const$),  freely falling in the cosmological  gravitational field,  so that their  hypersurfaces of constant cosmic time  are maximally symmetric. We also discuss how this class of cosmological observers relates to 
the standard R\&W ones. 

It is a compelling philosophical argument, and,  at the same time, a well established observational fact that the universe looks uniform to any R\&W observer, {\it i.e.}, to observers  that  are freely falling 
in the cosmic gravitational field (geodesic motion) and that are at rest (comoving) with the surrounding cosmic fluid  \cite{BE,UCE,cla,mbb,vmc}. It is worth noticing that, as shown in  appendix A, 
the most general class of geodesic and comoving cosmological observers ($t,x^{i}$) for which the hypersurfaces of constant time are flat and maximally symmetric 
can be described via the infinitesimal line element 
\begin{align}
ds^2 = dt^2 + 2\:\! q_i\:\! dt\:\! dx^i - a(t)^2 \delta_{ij}\:\! dx^i dx^j, 
\label{eq:metric_mixing}
\end{align}
where $q_i$ are constants, $a(t)$ is an arbitrary function of the cosmic time $t$,  and $\delta_{ij}$ is the Kronecker symbol. As a matter of fact, this metric has (at least) six spatial isometries,
 three translations $\bs{T}_ i$  and three rotations $\bs{R}_ i$, represented by the infinitesimal  Killing vectors 
\begin{align}
\bs{T}_ i = \frac{\pd}{\pd x^i} \;,\quad  \bs{R}_i = \epsilon_{ij}^{\;\;\;k}\left (x^j - q^j \!\int\!\! \frac{dt}{a(t)^2}\right )\! \frac{\pd}{\pd x^k} \quad {\rm for} \quad i=1,2,3
\label{eq:kil3}
\end{align}
where $q^i = \delta^{ij} q_j$ and where $\epsilon_{ij}^{\;\;\;k}$ is the Levi-Civita symbol\footnote{$\epsilon_{ij}^{\;\;\;k}$  equal 1 if $(i,j,k)$ is an even permutation of $(1,2,3)$, $-1$ if $(i,j,k)$ is an odd permutation of $(1,2,3)$ and 0 for repeated indices.}. 

In the local,  comoving,  geodesic coordinate system ${\it x^\mu}$,  cosmic  time is tilted,  it is  non-orthogonal
to spatial hypersurfaces.   For such observers, indeed, the time of flight  of photons along null geodesics is direction dependent. Consider a light cone with vertex $A \in \mathcal{M}$ and a light signal over some infinitesimal path.  The signal must satisfy $ds^2=0$, and  thus, by assuming for the sake of clarity that only $q_3\equiv q \ne 0$, 
\begin{align}
dt_{\pm} = - q dz  \pm \sqrt{q^2 dz^2 + a^2 (dx^2+dy^2+dz^2)}, 
\label{eq:asym_time}
\end{align}
\noindent where the plus sign corresponds to the future light cone at $A$, the negative
sign to the past light cone. There are two solutions for $dt$, corresponding to the two ways  
of taking the path; and both solutions are time dependent, therefore the 
metric  is not static \cite{Rindler}. Note that  the asymmetry in the propagation of light, {\it i.e.}, the time irreversibility,  
is induced by the preferred direction in time (and not in space) brought 
about by the time-space cross term $q$. It is thus worth mentioning, to avoid any possible misunderstanding,  
that  clock synchronisation between the fundamental observers, is still possible  in the frame $q \neq 0$,  and it can be operationally
carried out, at least in principle,  via light exchanges (e.g. \cite{cargiu}).\footnote{Time asymmetry  is exactly what we experience in the universe
under real `observing  conditions'. For example, consider a   flat R\&W metric $ds^2=d\tau^2-a^2 (d\mathrm{x}^2+d\mathrm{y}^2)$,  and look at the physical distance as a new spatial
coordinate  
\begin{eqnarray}
\bar{\tau} & =  & \tau  \nonumber \\
\mathrm{\bar{x}}& = & a \, \mathrm{x} \nonumber  \\
\mathrm{\bar{y}} & = & a \, \mathrm{y} \nonumber 
\end{eqnarray}

\noindent then, in the {\it ``physical"} reference frame, the metric element  reads 

\begin{equation}
ds^2=\left [ 1-\mathcal{H}^2 (\mathrm{\bar{x}}^2+\mathrm{\bar{y}}^2) \right]  d\bar{\tau}^2 +2\mathcal{H} (d\mathrm{\bar{x}}+d\mathrm{\bar{y}}) d\bar{\tau} -( d\mathrm{\bar{x}}^2 + d\mathrm{\bar{y}}^2), \nonumber
\end{equation}
where $\mathcal{H}$ is the Hubble parameter of the R\&W observers.}

It is worth clarifying here the relation between the `tilted-time' cosmological observers and the standard R\&W ones ($\tau, y^i$), those characterised by  the line element 
\begin{align}
ds^2 = d\tau^2 - a(\tau)^2 \delta_{ij}\:\! dy^i dy^j, 
\label{eq:rw}
\end{align}
where $\tau$ is the cosmic time of standard cosmology. Given that the line elements  \eqref{eq:metric_mixing}  and \eqref{eq:rw} are isometric,   there is  a diffeomorphism that  maps the  coordinates  $x^{\alpha}$ into the coordinates $y^{\alpha}$.  To this purpose we consider a class of freely falling particles that, at present  time $\tau_0$,  move with a given velocity $v_0$ with respect to the R\&W observers. Without lack of generality, we can orient both the  ${x^3} \equiv z $ and $y^3$  axes  in such a way that they are collinear to  the direction of this velocity.  The geodesic motion of these particles (see Eqs. \eqref{eq:12} in Appendix B)  is given,  in terms of the R\&W $y^i$ coordinates, by
\begin{eqnarray}
y^1(\tau) & =& \mathrm{x},  \label{sys1} \\
y^2(\tau) & =& \mathrm{y},   \label{sys2} \\
y^3(\tau) & =& \mathrm{z} \pm \int_{\tau_0}^{\tau} \frac{dt}{ a(t) \sqrt{1+ q^{-2} a^2(t)}}, 
\label{sys3}
\end{eqnarray}
where  
\begin{equation}
q \equiv \sqrt{ \frac{v_0^2}{1- v_0^2}}
\label{qdefi}
\end{equation}
\noindent and where the constant values $\mathrm{x}, \mathrm{y}, \mathrm{z}$ define the cosmological observers  of the standard cosmological model, {\it i. e.},   the geodesic observers with null velocity ($q=v_0=0$) at present time.  We now show that the comoving observers $x^i=const$ in the tilted time coordinate system  are those in motion with constant velocity $v_0$ with respect to R\&W observers, 
those that co-moves with the freely falling boosted particles.  Indeed, by setting $z=y^3(\tau)$ and by taking the differential,  we obtain  that the  infinitesimal coordinate transformation 
\begin{eqnarray}
d\mathrm{z}& = & dz + \frac{q}{a^2}dt \label{ttran}  \\
d\tau & = & dt \sqrt{1+\frac{q^2}{a^2}} \label{xtran} 
\end{eqnarray} 
maps  \eqref{eq:rw} into the infinitesimal line element 
\begin{equation}
ds^2 = dt^2 +  2\:\! q\:\! dt \:\! dz - a(t)^2\big(dx^2+dy^2+dz^2\big)\,, \qquad q={\rm const.}
\label{eq:line_element_gtz=q}
\end{equation}
Since we can always orient  the coordinate system $x^{\mu}$ such that two of the $q_i$ are made zero, from now on, and without lack of generality,  we will set  $q_1=q_2=0$ and $q_3=q$. 

The first thing worth  noticing is that, as anticipated, and  now made explicit by Eq. \eqref{qdefi},  the class of boosted comoving observers contains the R\&W one in the limiting case  $v_0=0$. More interestingly,   equation \eqref{xtran} shows that the hypersurfaces of constant time for the boosted observers are the very same hypersurfaces of the R\&W observer. This explains why  spatial hypersurfaces  
are maximally symmetric, in other terms  why the cosmological principle  holds also for this class of observers. 
The cosmological  time of boosted observers, instead, ticks at a different, slower,  rate with respect to the cosmic  time of the R\&W observers. 
We immediately  deduce  that  cosmic time  {\it is not univocally
defined} simply by requiring the clocks to be freely falling in the cosmological gravitational field. In other terms {\it one cannot transform  the coordinates $t$ into $\tau$ and require both 
to measure cosmic  time}. An additional physical structure is required if we are to specify cosmic time in an unambiguous, universal way. 
It is true that the intrinsic geometry of the cosmological manifold is invariant, {\it i.e.}, a coordinate transformation  does not change  the isometries of the metric, but is equally fundamental to 
recognise that  one cannot define two different cosmological observers (in the sense of Eq. \eqref{sys3}) and impose that they are both comoving with the same cosmological fluid. 
The observers \eqref{eq:metric_mixing} and \eqref{eq:rw}, although related by a diffeomorphism, do not describe the same physical system owing to the fact that 
they are both required to have a null spatial  velocity  ($\bs{u}_t \propto \partial_t$ and $\bs{u}_{\tau} \propto \partial_{\tau}$ respectively)   with respect to the {\it substratum}. 
One therefore needs to explore, using the EFE,  which cosmic fluid, if any,  is uniformly comoving  with the new reference frame.  
The  matter content of the universe will then eventually define the geometry of the cosmic spacetime which is compatible with this new hypothetical class of observers.

To get a grasp on the different physical content of the world model \eqref{eq:metric_mixing} with respect to the standard one (cf. Eq. \eqref{eq:rw}),  consider a time-like velocity field describing the motion of 
fluid  elements $dx^{\alpha}/d\lambda = u^{\alpha}(x^{\beta})$  (where $\lambda$ is an affine parameter, where  we assume that the world line of a fluid passes through every point $x^{\beta}$ of a certain region in spacetime, and, also,  that the velocity field of the fluid is differentiable in this whole region)  and assume, further,   that  the  velocity vector is  normalised to unity
\begin{align}
||\bs{u}|| = \sqrt{u^{\mu}\:\! u^{\nu} g_{\mu\nu}} = 1. \,
\end{align}
Its  covariant derivative can be decomposed in terms of  its trace ($\Theta$), its trace-free symmetric part ($\sigma^{\mu}_{\;\;\nu}$), and its (trace-free) antisymmetric part ($\omega^{\mu}_{\;\;\nu}$) as
 \begin{align}
u^{\mu}_{\;\;;\nu} =  \tfrac{1}{3} \:\! \Theta \:\! h^{\mu}_{\;\;\nu} + \sigma^{\mu}_{\;\;\nu} + \omega^{\mu}_{\;\;\nu},\,
\end{align}
where $h_{\mu \nu}=g_{\mu \nu }-u_{\mu}u_{\nu}$ is the projection tensor on hypersurfaces orthogonal to $u^{\alpha}$.  
The three  coordinate invariant scalar quantities defined as 
\begin{align}
\Theta &= u^{\alpha}_{\;\;;\alpha}  \quad \text{(expansion),}  \\
\sigma &= \sqrt{\sigma^{\mu}_{\;\;\nu} \, \sigma^{\nu}_{\;\;\mu}} \quad \text{(shear),} \\
\omega &= \sqrt{\omega^{\mu}_{\;\;\nu} \, \omega^{\nu}_{\;\;\mu}} \quad \text{(vorticity),}
\end{align}
characterise, univocally,  the evolution of the velocity field and a straightforward evaluation of these quantities shows that if the fluid is comoving (${\bf u}=(1,0,0,0)$) with the observers 
defined via Eq.  \eqref{eq:line_element_gtz=q} then 
\begin{align}
\Theta &= 3\:\! \frac{\dot{a}}{a}\,, \\
\sigma &= \sqrt{\frac{3}{2}} \frac{\dot{a}}{a} \frac{q^2}{a^2 +  q^2}\,,  \label{eq:sher}\\
\omega &= 0.
\end{align}
Notice that the shear scalar is different from zero only if the expansion scalar is different from zero and the time-space component $q$ is also non-null. 
We note, incidentally, that by using the Raychaudhuri equation,  the trace of the tidal tensor is
\begin{align}
E[u^{\mu}]^{\alpha}_{\;\;\alpha} & =  \Theta_{;\alpha} u^{\alpha} + \tfrac{1}{3}\Theta^2 + \sigma^2 - \omega^2 \\
& =  - 3 \frac{\ddot{a}}{a} + \Delta \Big(\frac{\ddot{a}}{a} - (1-\Delta) \frac{\dot{a}^2}{a^2}\Big), \,
\end{align}
where we define $\Delta \equiv q^2 / (a^2+q^2)$. In the special case where $\ddot{a}/a = (1-\Delta) (\dot{a}/a)^2$, we recover the FLRW result $E(u^{\mu})^{\alpha}_{\;\;\alpha} = -3 \:\! \ddot{a}/a$. 

The centrality of the assumption $q=0$ in singling out the standard cosmological observers   can now be understood in terms of a necessary and sufficient condition for an observer to be in the R\&W class.  
A cosmological observer is of the R\&W type if and only if  it obeys EFE with a {\it perfect fluid} source that has zero vorticity, shear and acceleration \cite{Krasinski}.\footnote{Note that, an equivalent definition   makes no reference
at all to Einstein's field equations: an observer is R\&W  iff  (1) the spacetime  admits a foliation into spacelike hypersurfaces of constant curvature, (2) the congruence of the observers worldlines are  orthogonal to the leaves of the foliation and shear-free geodesics, and  (3) the expansion scalar of the geodesic congruence has its gradient tangent to the geodesics.}
Equation \eqref{eq:sher} shows now that  the boosted observer  implements these properties except for the shear-free condition. 
Although the new coordinates are defined via a diffeomorphism (cf. Eqs. \eqref{ttran} and \eqref{xtran}),  the shear scalar is not equal to the value calculated using R\&W coordinates. 
Indeed we have imposed that the velocity of the fluid is null in the boosted coordinates, {\it i.e.},  it is  not fixed by  the coordinate transformation itself. 
Note, however, that the R\&W value  can be recovered in the asymptotic limit in which $a(t)>>q$. In other terms it is  the dynamical evolution of the scale factor that eventually drives the general metric \eqref{eq:metric_shear} into the R\&W special limit. 
That this effectively happens,  at large cosmic times,  will be seen in \S 3.2.

In what follows we shall be investigating the physical consequences of  relaxing the assumption that the cosmic  time measured by a class of comoving observes
is orthogonal to space-like  hypersurfaces.  There are several reasons that make it  natural to explore the aftermath of 
abandoning this  assumption. The simplest is that, as we will see in the next sections,  this  class of comoving observers has an interesting level  of generality, encompassing the R\&W subclass in
a given well-defined limit.  The most  compelling, however,  is that any increase in  generality  which can be brought about without introducing new hypotheses,  but 
only by removing  previous restrictions,  will be of advantage in shedding light on the intrinsic nature of the  cosmic spacetime.

\section{Dynamics}
In this section we explore which cosmic component, if any, satisfies the cosmological principle  in the reference frame of the boosted comoving geodesic observers.
We do this by computing the EFE for the metric  \eqref{eq:line_element_gtz=q} and by assuming a general relativistic imperfect fluid stress energy tensor. 

\subsection{Einstein's Field Equations}

A vigorous research program is nowdays conducted to  assess whether the predictions of the EFE can be safely extrapolated on cosmological scales of the order of the 
Gigaparsec \cite{CFPS,PV,BNP,psm,SBM,bbmv,RHFS,TA}.   Current evidences seem to suggest that this is indeed the case only if the EFE is complemented by an additional term, the cosmological constant $\Lambda$ \cite{per,riess,mb,koma,bm,pla}. Since our goal is to highlight the dynamical emergence of  cosmic components that  behave as an effective cosmological constant (in the sense that their 
equation of state is $p=-\rho$), we will assume,   right from the beginning,   that the EFE  contain  no additional $\Lambda$ term.  The  Einstein tensor   is  
\begin{equation}
G^{\mu}_{\;\;\nu} = R^{\mu}_{\;\;\nu} - \frac{1}{2} R^{\alpha}_{\;\;\alpha} \delta^{\mu}_{\;\;\nu}
\end{equation}
and it has the following non-vanishing components
\begin{align}
G^{0}_{\;\;\,0} &= 3(1-\Delta)H^2 \,, \\
G^{3}_{\;\;\,0} &= -2\frac{q}{a^2}(1-\Delta) \big(\dot{H}+\Delta H^2\big) \,,  \\
G^{i}_{\;\;i} &= (1-\Delta)\big( 2\dot{H}+(3+2\Delta)H^2\big) \,\,\,\,\,  \text{(no summation on repeated indeces).}
\label{eq:einstein_tensor}
\end{align}
Note that $G^{0}_{\;\;i}=0$, and in particular $G^{0}_{\;\;3}=0$,
%
Following \cite{Weinberg1972}, the stress-energy tensor of a  general relativistic imperfect fluid  can be modelled as 
\begin{align}
T_{\mu\nu} = & \rho(t)\:\! u_{\mu}\:\! u_{\nu} - p(t)\;\! h_{\mu\nu}  \nonumber \\
& - \eta(t)\:\! h^{\alpha}_{\;\;\mu}\:\! h^{\beta}_{\;\;\nu} \:\!\sigma_{\alpha\beta} - \zeta(t)\:\! \Theta \:\! h_{\mu\nu} - \chi(t) \big(h^{\alpha}_{\;\;\mu}\:\! u_{\nu} + h^{\alpha}_{\;\;\mu} \:\! u_{\nu}\big)\big( \mathrm{T}_{;\alpha} + \mathrm{T}\:\! \dot{u}_{\alpha}\big)
\end{align}
where $h_{\mu \nu}$ is the projection tensor, $\rho=\rho(t)$ and $p=p(t)$ are respectively energy density and isotropic pressure of the fluid, and $\eta(t)$, $\zeta(t)$ and $\chi(t)$ can be interpreted as {\it shear viscosity}, {\it bulk viscosity} and {\it heat conduction}. The first line contains  perfect fluid terms, while the dissipative terms are  in the second line. Due to homogeneity and isotropy, we allow all these 5 functions to depend only on time. This should hold for  the temperature $\mathrm{T}$ defined such that the energy density $\rho(\mathrm{T},n)$ is equal to the comoving energy density $T_{\alpha\beta}u^{\alpha}u^{\beta}$ at thermal equilibrium. Writing explicitly the temperature as $\mathrm{T}=\mathrm{T}(t)$ induces the tensor multiplying $\chi(t)$ to vanish. If we additionally impose that the fluid is  comoving (${\bf u}=(1,0,0,0)$) with the tilted-time observers  $x^{\mu}$,  we are left with the following components
\begin{align}
T^{0}_{\;\;\;0} &= \rho(t) \nonumber \\
T^{0}_{\;\;\;3} &= q \big(\rho(t) + p(t)  -\tfrac{2}{3}\Delta H\:\!\eta(t) + 3(1-\tfrac{1}{3} \Delta)H\:\! \zeta(t)\big)  \nonumber \\
T^{1}_{\;\;\; 1} &= T^{2}_{\;\;2} = -p(t) -\tfrac{1}{3}\Delta H\:\! \eta(t)-(3-\Delta)H\:\! \zeta(t) \nonumber \\
T^{3}_{\;\;\; 3} &= -p(t) - \tfrac{2}{3}\Delta H\:\! \eta(t) - (3-5\Delta) H \:\! \zeta(t). 
\end{align}
Note, incidentally, that  $T^{3}_{\;\;\;0}=0$. By comparing the  above equations with \eqref{eq:einstein_tensor}, we see that the components $G^{i}_{\;\;i}$ are the same whereas $T^{i}_{\;\;i}$ are not. By 
enforcing the equality of the components $T^{i}_{\;\;i}$, a stringent constraint between shear and bulk viscosity
\begin{align}
\eta(t) = 12 \zeta(t)
\end{align} 
results.  As a consequence, we obtain  the following four {\it Friedmann-like} equations 
\begin{align}
3H^2(1-\Delta) & = 8\pi G\rho,  \\
-2\tfrac{q}{a^2}(1-\Delta)\big(\dot{H} + \Delta H^2\big) &= 0,  \\
0 &= 8\pi G q\big(\rho + p + 3(1-3\Delta)H \zeta \big),  \\
(1-\Delta)\big(2\dot{H} + (3+2\Delta)H^2\big) &= -8\pi G \big(p +3 (1+\Delta) H \zeta \big).
\label{eq:Friedmann_4} 
\end{align}

Moreover,  the  conservation equation, 
$T^{\mu}_{\;\;\;\nu;\mu} = 0$, yields the additional, non-independent,  equations (for $q \neq 0$) 
\begin{align}
\dot{\rho} &= - (3-\Delta)H\big[\rho + p + 3(1+\Delta)H \zeta \big] \label{eq:dotrho}\\
\dot{p} &= 6\Delta(1+3\Delta)H^2\zeta -3(1-3\Delta)(\dot{H}\zeta + H\dot{\zeta}). \label{eq:dotp}
\end{align}
We are thus led to the conclusion that if viscosity is zero then the pressure (if any) has to be constant in time.

\subsection{Solutions}
If we set $q=0$ ($\Delta = 0$) in equation \eqref{eq:Friedmann_4} we obtain the usual Friedmann equations for an imperfect fluid:
\begin{align}
3H^2 &= 8\pi G \rho \\
2\dot{H} + 3H^2 &= -8\pi G (p +3 H \zeta).
\end{align}
Notice that for the case of the Friedmann equations, the equation of state parameter $w=p/\rho$ is completely unconstrained.
If $q\neq 0$ ($\Delta \neq 0$), instead, we obtain
\begin{align}
3H^2(1-\Delta) & = 8\pi G\rho, \,  \label{eq:Fr_1} \\
\dot{H}+\Delta H^2  &= 0,  \, \label{eq:Fr_2}\\
p & = -\rho,  \,   \label{eq:Fr_3} \\
\zeta &= 0. \label{eq:Fr_4}
\end{align}
that is, no real  fluids, only perfect ones, are solutions of the EFE. Additionally,  their equation of state is constrained and violates the 
strong energy condition, {\it i.e.}, the requirement $\rho+3 p \ge 0$. 

By inserting eqs. \eqref{eq:Fr_3} and \eqref{eq:Fr_4} into  \eqref{eq:dotrho} we deduce that the energy density $\rho$ it is constant in time. The  
general solution of equation (\ref{eq:Fr_1}) is thus
\begin{equation}
a(t) =  e^{\hL (t-t_{0})} - \frac{q^2}{4} e^{-\hL (t-t_{0})}, 
\label{eq:a_solution}
\end{equation}
where $t_{0}$ is an arbitrary integration constant and where we have expressed the constant density component   as the vacuum energy seen by a 
R\&W observer
\begin{equation}
\rho=const \equiv  \frac{3 \hL^2}{8\pi G}.
\label{eq:rholambda}
\end{equation}
Indeed, by definition, any freely falling observer sees the same vacuum (which is Lorentz invariant).  Note that, unlike $H$,  the Hubble parameter  $\hL$  for  a R\&W observer is constant. 

If $q \ne 0$, then, in  the finite past, the  geodesic worldlines intersect at a point, {\it i. e.},  the scale factor vanishes. By imposing that this happens for $t=0$
we can fix the arbitrary integration constant 
 \begin{equation}
 t_{\rm 0} =- \frac{1}{\hL } \ln \frac{q}{2},  
\end{equation}
and we can recast  the solution \eqref{eq:a_solution} in the form
\begin{equation}
a(t) = q \sinh \left (\hL t\right).
\label{eq:a_solution_rescaled}
\end{equation}

Note that even though $H$ is not a constant, we are dealing with a de Sitter spacetime. In fact, evaluating the Ricci scalar, we find
\begin{align}
R^{\alpha}_{\;\;\alpha} =  - 6 (1-\Delta)\big(\dot{H} + (2+\Delta)H^2\big)= -12(1-\Delta)H^2 = -32\pi G \rho, 
\label{rc}
\end{align}
which is manifestly constant.   That this  solution represents a 4D de Sitter universe is also made evident  by the fact that the spacetime has maximal symmetry, that is 10 Killing vectors.
Indeed by solving the Killing equations (\eqref{eq:Killing} in Appendix A)  using  the metric  \eqref{eq:metric_mixing}  and the scale factor \eqref{eq:a_solution_rescaled} 
we find, on top of the spatial translations $\bs{T}_i$ and rotations $\bs{R}_i$ (see cf. Eq. \eqref{eq:kil3}), also the time translation isometry $\bs{T}_0$  and three Lorentz isometries (boosts)
$\bs{B}_i $. Explicitly,
\begin{align}
\bs{T}_0 = & \;  \hL ^{-1} \tanh(\hL t) \:\! \pd _t - \big( x^k + q^k (q^j q_j)^{-2} \hL^{-1} \tanh (\hL t) \big) \pd _k \, ,    \label{eq:Tt_dS_tilt_3} \\
\bs{T}_i = & \; \pd _i \, ,  \label{eq:Ti_dS_tilt_3} \\
\bs{B}_i = & \; \hL^{-1} \big( x_i \tanh(\hL t)  + q_i (q^j q_j)^{-2} \hL^{-1} \big)\!\: \pd_t  \nonumber \\
           & \; - \big(x_i + q_i \:\! (q^j q_j)^{-2} \hL^{-1} \coth(\hL t) \big) \big(x^k + q^k (q^j q_j)^{-2}\hL^{-1} \tanh(\hL t) \big)\:\! \pd _k  \nonumber \\
            & \;  + \tfrac{1}{2} \Big[ \hL^{-2} (q^j q_j)^{-2} + x_k \big(x^k + 2\!\: q^k (q^j q_j)^{-2} \hL^{-1} \tanh(\hL t) \big) \Big] \pd_i  \, ,  \label{eq:Li_dS_tilt_3} \\
\bs{R}_i = & \;  \epsilon_{ij}^{\;\;\;k} \big(x^j + q^j (q^l q_l)^{-2} \hL^{-1} \coth(\hL t) \big)\:\! \pd _k  \, , \label{eq:Ri_dS_tilt_3}
\end{align}
where the Kronecker symbol  $\delta_{ij}$ is used to rise/lower spatial indexes.  These Killing vectors  obey the  non vanishing de Sitter commutation relations
\begin{align}
& [\bs{T}_0,\bs{T}_i]=\bs{T}_i \,, \quad [\bs{T}_0,\bs{B}_i] = -\bs{B}_i\,, \quad [\bs{T}_i,\bs{B}_j] = \delta_{ij} \bs{T}_0 + \epsilon_{ij}^{\;\;\;k} \bs{R}_k \\
& [\bs{T}_i,\bs{R}_j] = - \epsilon_{ij}^{\;\;\;k} \bs{T}_k \,, \quad [\bs{B}_i,\bs{R}_j] = -\epsilon_{ij}^{\;\;\;k} \bs{B}_k \,,\quad [\bs{R}_i,\bs{R}_j] = -\epsilon_{ij}^{\;\;\;k} \bs{R}_k
\end{align}
{\it i.e.}, the  commutations of the  $O(4,1)$ group in the basis defined by our coordinates.  

In  summary, if we describe gravity using the EFE without the cosmological constant,  the only model of the universe that looks uniform    
to freely falling, comoving, boosted observers  is de Sitter. In other terms,  
the universe cannot contain a geodesic fluid,  comoving with the observers,  other than  a perfect one  with an equation of state parameter  $w=p/\rho = -1$.
This analysis complements the standard  dynamical derivation of de Sitter spacetime as the vacuum  solution of the EFE {\it augmented} by the cosmological 
constant in a universe which looks uniform to R\&W observers.  It highlights that  a component with an effective equation of state  $p=-\rho$ (be it a cosmological constant or dark energy)  is not only a mathematical  option,   it is, instead, and under very general conditions, a necessary condition imposed  by  requiring the validity of the cosmological principle for a general class of freely falling, 
comoving, observers.  In other terms, such  an `exotic' equation of state   naturally emerges as  an essential  and not ancillary  ingredient of  the standard model of the universe  if space is to show  translational and rotational symmetries.

Besides  offering new insights  into the dynamical emergence  of de Sitter space times in cosmology,  
this result also highlights, in some sense,   the intrinsic non-Machian nature of the general relativistic theory of gravity.
The issue whether general relativity realises Machian ideas has always been a controversial one, especially because, in spite of much discussions and debates,  it has never been entirely clear what the 
Mach principle is. With this caveat in mind, it is nonetheless interesting to note  that while the theory allows for the possibility of the boosted motion of 
observers within the gravitational field generated by  a uniform distribution of matter, a global boost of a uniform distribution of matter cannot generate  that very same gravitational field.
In other terms, from this arguments it seems that  there is no such principle as the relativity of inertia embedded in the EFE.   

\subsection{Slicing the de Sitter spacetime}

Equation \eqref{eq:a_solution} shows that the usual  de Sitter expansion law (in flat R\&W coordinates) is recovered in the special case  $q=0$,
More interestingly,  also at large  times the scale factor  expands  as  in the flat R\&W coordinates.  Indeed, as soon as $t \gg  \hL ^{-1}$, $q$ becomes subdominant with respect to  $a(t)$, the shear scalar vanishes  (cf. Eq. \eqref{eq:sher}),  and the boosted geodesic observer \eqref{eq:line_element_gtz=q} asymptotically converges to the  flat R\&W one, {\it i.e.}, the boosted  and the R\&W frames are essentially undistinguishable.

The coordinate transformation that maps the comoving coordinates of the boosted observer $(t,x,y,z)$ into the flat R\&W coordinates of a de Sitter space time are obtained by embedding a 4D 
hyperboloid in a 5D Minkowskian embedding space $(Y_0,Y_1,Y_2,Y_3,Y_4)$ with coordinate transformation
\begin{align}
Y_0 &= \hL^{-1} \left [  \left (1 + \tfrac{1}{2}\big(x^2+y^2+ z^2\big) \right)  \sinh \left (\gamma\,t \right ) +z \cosh \left ( k\, t \right ) \right ] \,, \label{eq:coord_x_0_tz0_tilted} \\
Y_1 &= - \hL ^{-1} \left [ z \cosh  \left (\gamma\, t \right )  + \tfrac{1}{2}\big(x^2+y^2+z^2\big) \sinh \left ( \gamma\, t \right ) \right ]\,, \label{eq:coord_x_1_tz0_tilted}\\
Y_2 &= \hL^{-1} \left[ \cosh \left ( \gamma\,  t \right ) + z \sinh \left ( \gamma\, t \right ) \right] \,, \label{eq:coord_x_2_tz0_tilted} \\
Y_3 &= \hL^{-1}\:\! x\:\! \sinh \left( \gamma\, t \right)\,, \label{eq:coord_x_3_tz0_tilted} \\
Y_4 &= \hL^{-1}\:\! y\:\! \sinh \left(\gamma\, t \right)\, ,  \label{eq:coord_x_4_tz0_tilted} 
\end{align}
where $\gamma$ is an arbitrary parameter.  One can verify that these coordinates define a 4-dimensional de Sitter hyperboloid,
\begin{align}
Y_0^2-Y_1^2-Y_2^2 -Y_3^2 - Y_4^2= -\mathcal{H}_{\Lambda}^{-2} \,,
\end{align}
where $Y_1=Y_2=Y_3=Y_4=0$ defines its axis of symmetry and  $\hL^{-1}$ is its radius  at $Y_0=0$. Also, one can verify that the coordinate transformation (\ref{eq:coord_x_0_tz0_tilted}) to (\ref{eq:coord_x_4_tz0_tilted}) induces the tilted time metric (\ref{eq:line_element_gtz=q}) on the hyperboloid,
\begin{align}
ds^2 &= dY_0^2 -dY_1^2 -dY_2^2 -dY_3^2-dY_4^2 \nonumber \\
  &= dt^2 +2\!\: q \:\! dt\!\: dz - q^2 \! \sinh^2 \left ( \hL t \right )\:\! \big(dx^2+dy^2+dz^2\big) ,
 \end{align}
once one identifies 

\begin{equation}
q=\hL^{-1}=\gamma^{-1}.
\end{equation}
This constraint between the constant vacuum  energy density (parameterised by $\hL$)  and the velocity  of the boosted observer (parameterised by $q$) follows from fixing the gauge freedom in 
\eqref{eq:a_solution},  {\it i.e.}, by defining the zero point of cosmic time as the specific moment at which the scale factor goes to zero. 

A visual picture of the de Sitter space foliation implied by our geodesic comoving coordinates can be obtained by setting $x=y=0$ in \eqref{eq:coord_x_0_tz0_tilted}-\eqref{eq:coord_x_4_tz0_tilted}.
The resulting 2D de Sitter hyperboloid, together with the coordinate grid, is shown in fig. \ref{fig:hyperboloid}. Noteworthily,   the tilted time coordinate system covers all of the hyperboloid, contrary to  the 
flat R\&W  coordinates which cover only half of it. As expected,  when the time  constraint $t \ge 0$ generated by the EFE is considered, the portion of de Sitter space covered by the two coordinate
systems is identical. Finally note that the coverage is not complete in 4D (only 85\% is charted).

\begin{figure}[htbp]
  \centering
  \begin{minipage}[b]{5 cm}
    \includegraphics[width=5cm]{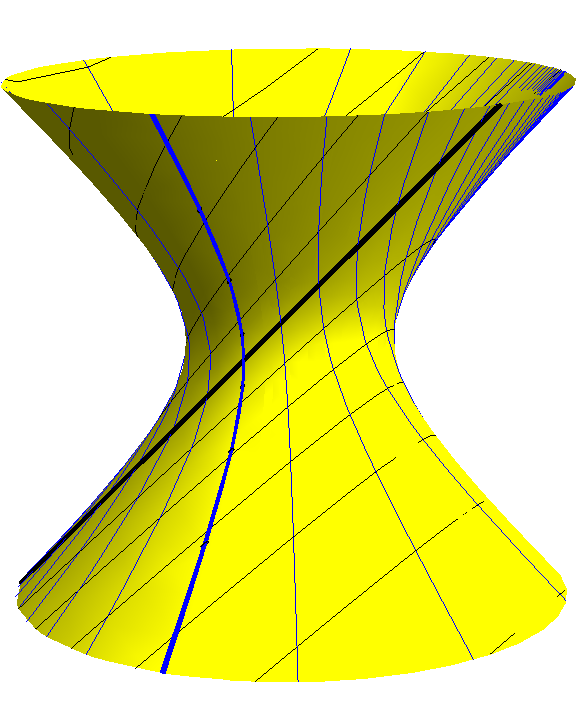} 
  \end{minipage}
  \hspace*{0.5cm}
  \begin{minipage}[b]{5 cm}
    \includegraphics[width=5cm]{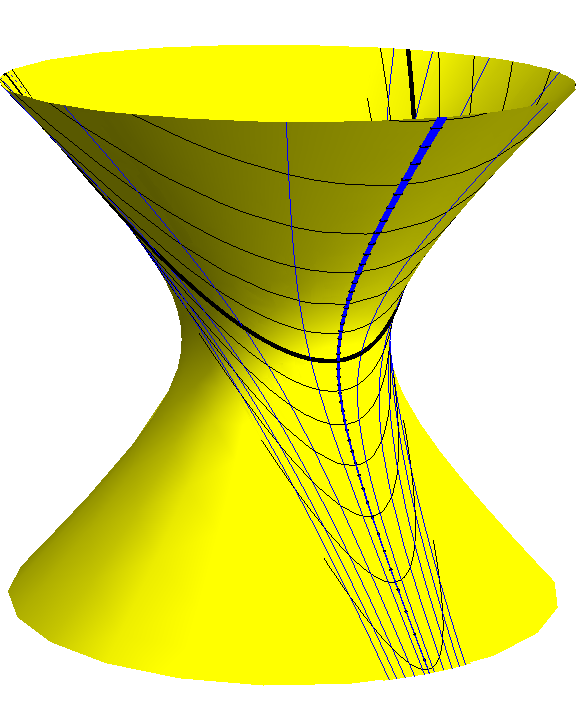}  
  \end{minipage}
       \caption{\textit{Left:} foliation of de Sitter spacetime with the tilted time coordinates. Blue lines represent comoving geodesics $z=const$, black lines represent hypersurfaces of constant cosmic time $t$. These coordinates cover the entire hyperboloid for $-\infty < t <\infty$ and $-\infty < z < \infty$. The thicker lines indicate the curves  $z=0$ and $t=0$
       \textit{Right:}  de Sitter foliation in the standard flat R\&W coordinates for $-\infty < \tau <\infty$ and $-\infty < \mathrm{z} < \infty$.  }
    \label{fig:hyperboloid}
\end{figure}

\section{Origin of time in de Sitter spacetime}

Although the scale factor $a(t)$ vanishes, there is no physical  singularity at $t=0$,  in the sense that the curvature invariant  \eqref{rc} is well defined. 
Instead the shear scalar (cf. Eq. \eqref{eq:sher})
\begin{align}
\sigma(t) = \sqrt{\frac{8}{3}} \frac{\hL}{\sinh \left ( 2\:\! \hL t \right ) }
\end{align}
diverges when $t \rightarrow 0$.   By exploiting the time dilation equation \eqref{ttran}, we  obtain the explicit relation between the cosmic  time measured by  boosted and R\&W observers

\begin{equation}
t=\hL^{-1} \left (e^{\hL \tau} +\sqrt{1+ e^{2 \hL \tau} }  \right ), 
\end{equation}
which makes manifest the mapping of the  R\&W epoch $\tau=-\infty$  into the coordinate $t=0$ of the tilted time coordinate system. Note also that  the two clocks tick at the same pace
at large cosmic times, as soon as $t \gg \mathcal{H}_{\Lambda}$.
 
We further discuss  this  specific  remapping of the cosmic time  by calculating the time evolution of a ghost condensate \cite{ghost} in  the boosted frame. 
The ghost condensate is a hypothetic fluid that might fill the universe  and that has the same effective equation of state of  the cosmological constant ($\rho = -p$). It can  hence drive de Sitter expansion of the universe. However, unlike a cosmological constant, it is a physical fluid which is unstable under scalar perturbations.  We consider  its
Lagrangian density $\mathcal{L}=-P$ where $P=-X +\frac{X^2}{M^4}$ and where $X=(1/2) g^{\mu \nu} \phi_{,\mu} \phi_{,\nu}$. By considering the kinetic term 
$X=M^4/2$,  the equation of state of the  effective perfect fluid associated to this cosmological component is $p=-\rho$, while the time evolution of the field is 
\begin{equation}
\phi(t)=\phi(t_i)\pm\frac{M^2}{\hL} \ln \left [ \frac{\sinh \left (  \hL t   \right ) }{\sinh \left(  \hL t_i \right )  } \right ].
\end{equation}
Note that the initial time $t_i$ cannot be set arbitrarily, but must satisfy  the constraint $t_i \ge 0$. Also note that the $\phi$-field evolves as in the 
standard flat R\&W coordinates ($\phi=\phi_i + M^2 (\tau-\tau_i)$) as soon as $\tau \sim t \gg \mathcal{H}_{\Lambda}^{-1}$.

\section{Discussion and conclusions}

An a-synchronous coordinate  system is considered in which the fundamental observer of the universe is not anymore the R\&W one -- the freely falling observer that is comoving with a 
perfect fluid  having zero  shear, vorticity and acceleration -- but a new one, that can still define a cosmic time and spatial hypersurfaces of
constant curvature.  Indeed,  the coordinate system 
\begin{align}
ds^2 = dt^2 + 2\:\! q_i\:\! dt\:\! dx^i - a(t)^2 \delta_{ij}\:\! dx^i\:\! dx^j
\label{final}
\end{align}
obtained by relaxing the shear-free assumption and by  considering as fundamental cosmological observers test particles that are  boosted with a given constant velocity with respect to the R\&W ones, 
is not, in principle,  in conflict with the requirements of the cosmological principle. Specifically,  a class of  freely falling comoving ($x^i=const$) observers,  with  proper time not orthogonal to the flat spatial hypersurfaces,  is identified for which the hypersurfaces of constant cosmic time  are maximally symmetric, {\it i. e.}  six independent isometries of the metric exist, three translations and three rotations. 

We find that the only fluid that can be naturally accommodated in such a universe, if the cosmological principle must hold, is  a perfect one  with an effective equation of state $p=-\rho$. 
In particular, when the Einstein's field  equations without $\Lambda$ are considered,  we find  that  the second time derivative of the scale factor of the metric is positive and equal to 
\begin{equation}
\ddot{a}=(8\pi G/3)a
\end{equation}
that is,  the metric expands at the same  accelerated  pace as  R\&W tests particles  in the cosmological model obtained by solving the EFE, augmented by $\Lambda$, in the vacuum.
We also demonstrate that  the resulting  1+3 spacetime is de Sitter, thus it possesses maximal symmetries.
This analysis contributes to shedding light on the dynamical emergence of de Sitter space times in physics, and more specifically, 
provides  an argument in favor of a natural emergence of the equation of state $w=-1$  in the context of  the standard cosmological model. Indeed, 
despite its effects being dynamical, such as accelerating the expansion of a set of geodesic comoving observers,  the very nature of this peculiar equation of state,
appears to be  intrinsically geometric, {\it i.e.},  associated to the requirement that space be maximally symmetric to a  general class of cosmological observers. 

We discuss several   interesting  cosmological properties of the tilted time  coordinate system. First we show  that it  covers de Sitter spacetimes in a complete way in 1+1 dimensions, and we  
demonstrate that  the  coordinate time, which also plays the role of cosmic time,   is finite in the past, a concept that we illustrate by showing that  the time evolution of  a specific dynamical scalar field with 
equation of state $p=-\rho$ -- the ghost condensate -- is bounded from below.  Finally,  at large times, when $a(t) \gg \sqrt{q_i q^i} $ the  geodesic slicing  of the de Sitter spacetime implied by this coordinate system  asymptotically converges to the  de Sitter slicing in standard flat  R\&W coordinates. That is,  the flat R\&W observer  is an attractor in the space of the general comoving observers  specified by the metric
element  \eqref{eq:metric_mixing},  irrespectively of the initial value of the constant parameters $q_i$. 

As a byproduct, we discuss how this novel way of looking at de Sitter spacetimes also sheds light on the anti-Machian character of the Einstein's theory of gravity. 
The fact that it is not possible to  generate a consistent gravitational field  by  boosting the uniform mass distribution of the universe,   is at variance with the somewhat common interpretation  of the Mach principle as a statement that inertia is a relative concept, {\it i.e.},  that boosting the sources of the gravitational field or the observers,  generates physically equivalent inertial effects. 
           
A way to generalise and improve the results presented in this paper is to figure out if there are  comoving cosmological observers  such that their proper time is non orthogonal to curved spatial hypersurfaces of maximal symmetry. An important aspect to explore is also  the perturbative instability of such a metric.   At a  more speculative level, one might examine whether  such  sheared fluid verifies  the conditions for  generating a  successful inflationary phase of the universe.  

\section*{Acknowledgments}

We acknowledge useful discussions with  L. Perenon, S. Lazzarini, U. Moschella, \"{O}.~Akarsu, J.~Bel, P.~Taxil and J.~M.~Virey. We  are grateful to 
R. Coqueraux and T.~Sch\"{u}cker  for critical reading of manuscript drafts and for providing valuable comments and suggestions.  
We  thank F. Piazza for  suggestions and inspiring discussions about the Mach principle.  
CM is grateful for support from specific project funding of the {\it Institut Universitaire de France} and of the Labex OCEVU.

\bibliographystyle{spr-chicago}      
\bibliography{example}   
\nocite{*}


\subsection{Appendix A.  The tilted time cosmological line element}

Among  the general class of reference frames in which the universe can be described and interpreted,  
one can   single out  as the fundamental ones those  associated to test particles that  are freely falling 
in the cosmic gravitational field, that is  in geodesic motion  $\bs{u}'= u'^{\mu} \tfrac{\partial}{\partial x}=0$,  where
\begin{align}
\bs{u} = \frac{dx^{\mu}}{d\lambda}\frac{\pd}{\pd x^{\mu}} = u^{\mu} \frac{\pd}{\pd x^{\mu}}
\label{vfield}
\end{align}
is the velocity vector,   $\lambda$ is some affine parameter, and  where  $' \equiv d/d\lambda$.
We assume that these fundamental cosmological observers are further characterised by a null spatial  velocity 
 $\bs{u} \propto \partial_t$,  {\it i.e.},, they are comoving with the coordinate system ($x^i=const$). 
 
The $(0,0)$ component of the metric  is  univocally determined by assuming that the coordinate time  of freely falling observers coincides with the proper time, {\it i.e.},  
\begin{equation}
g_{00}=1.
\end{equation}
This same  condition  allows to constrain the functional dependence of the mixed components $(0,i)$ of the metric. Given the geodesic equations of motion
\begin{equation}
u'^{\mu} = \frac{d^2x^{\mu}}{d \lambda^2} + \Gamma^{\mu}_{\,\,\,\alpha\beta} \frac{dx^{\alpha}}{d\lambda}  \frac{dx^{\beta}}{d\lambda} = 0 \;
\label{eq:geodesic_equation}
\end{equation}
\noindent  one obtains,  for  comoving  observers,  $\Gamma^{\mu}_{\,\,\,00}=0$ or, equivalently,
\begin{equation}
\frac{\pd g_{0i}}{\pd t} = 0 \qquad i=1,2,3
\end{equation}
\noindent that is,  the cross-terms between the time component and the space  components can be at most generic functions of space positions.

Any non-zero 4-displacement with $dt=0$ takes place  in what is generally called a spatial hypersurface $\sigma_t$. 
Standard cosmology follows from the assumption that flat hypersurfaces have 
six isometries, three translations and three rotations, represented by the infinitesimal operators
\begin{equation}
^{(3)}\bs{T}_i = \frac{\pd}{\pd x^i}\,, \qquad ^{(3)} \! \bs{R}_i = \epsilon_{ij}^{\;\;\;k} x^j \frac{\pd}{\pd x^k}\,, \qquad i=1,2,3.
\end{equation}
These Killing vectors form a basis of the space of solutions of the Killing equation for the metric  $ \delta_{ij}\:\! dx^idx^j$.
The spatial isometries in a 4-dimensional spacetime with flat space sections will therefore be of the general type 
\begin{align}
\bs{T}_i = F_{\underline{i}}(t)\,\frac{\pd}{\pd x^i}\,, \qquad \bs{R}_i =  \epsilon_{ij}^{\;\;\;k} \, (G_{\underline {k j}}(t) x^j + H_{\underline{k}}^j(t) )\frac{\pd}{\pd x^k}\,, \qquad i=1,2,3 
\label{eq:Killing_Ansatz}
\end{align}
where $F_{\underline{i}}(t)$, $G_{\underline{k j}}(t)$  and  $H_{\underline{i}}^j(t)$ are, respectively,   3, 6 and 6  {\it a-priori} arbitrary functions of time  and where 
the summation convention is disabled on underscored indices.

Since  equations \eqref{eq:Killing_Ansatz} obey the Euclidian commutation relations
\begin{align}
& [\bs{T}_i,\bs{T}_j] = 0 \,, \label{eq:commutator_1} \\
& [\bs{T}_i,\bs{R}_j] = - \epsilon_{ij}^{\;\;\;k} \bs{T}_k \,, \label{eq:commutator_2} \\
& [\bs{R}_i,\bs{R}_j] = - \epsilon_{ij}^{\;\;\;k} \bs{R}_k \,,  \label{eq:commutator_3}
\end{align}
Eq. \eqref{eq:commutator_1} is trivially satisfied. Equation \eqref{eq:commutator_2}, instead,  imposes $F_{\underline{i}}(t)$  to be an isotropic function, $F(t)$, while  from equation \eqref{eq:commutator_3} we deduce     $G_{\underline{k j}}(t)=1$ and $H_{\underline{i}}^j(t)   =H^j(t)$.  

We can  further specify the components of the metric   by requiring them to be solutions of the  $6 \times 10 = 60$  Killing equations 
\begin{align}
\xi^{\mu} \frac{\pd g_{\alpha\beta}}{\pd x^{\mu}} + \frac{\pd \xi^{\mu}}{\pd x^{\alpha}} \,g_{\mu\beta}+ \frac{\pd \xi^{\mu}}{\pd x^{\beta}}\, g_{\alpha\mu} = 0
\label{eq:Killing}
\end{align}
where $\xi^{\mu}(x)=\{(\bs{T}_1)^{\mu},...,(\bs{R}_3)^{\mu}\}$. We solve these partial differential  equations by making the ansatz
\begin{align}
 ds^2 = dt^2 + 2\:\!  g_{0i}(x^j) dt\:\! dx^i  - a(t)^2 \left ( dx^2+dy^2+dz^2  \right )
\label{eq:metric_shear}
\end{align}

\noindent The  translations $\bs{T}_i$ yield the following non-trivial equations   
\begin{align}
 2\:\! g_{0i}(x^j) \frac{dF}{dt} & = 0\,, \\
-a(t)^2 \frac{dF}{dt} + F(t) \frac{d g_{0i}}{dx^j} &= 0\,, \quad i=j\\
F(t) \:\! \frac{d g_{0i}}{dx^j} & = 0\,, \quad i\neq j
\end{align}
where $i={1,2,3}$ and summation over indices is again omitted.
Solving these equations we find that $F(t)=const$ and $g_{0i}(x^j) = const$. 
We can set without loss of generality $F(t)=1$ and $g_{0i}(x^j)=q_i$. By solving the Killing equations for the rotations $\bs{R}_i$ we obtain the equations
\begin{align}
\delta^{ij} q_j  + a(t)^2 \frac{dH^i}{dt} &= 0 \,,\quad i=1,2,3 \label{eq:R_2}
\end{align}
and therefore
\begin{align}
H^i(t) = - \int\! \frac{q^i}{a(t)^2} dt \,,
\end{align} 
where we define $q^i = \delta^{ij} q_j$. This implies  the following form for the Killing vectors 
\begin{align}
\bs{T}_ i = \frac{\pd}{\pd x^i} \;,\quad  \bs{R}_i = \epsilon_{ij}^{\;\;\;k}\left (x^j - q^j \!\int\!\! \frac{dt}{a(t)^2}\right )\! \frac{\pd}{\pd x^k} \quad {\rm for} \quad i=1,2,3.
\end{align}

As a consequence, the general  line element given  by Eq. \eqref{eq:metric_mixing}  verifies the cosmological principle.

%

\section{Geodesics motion of freely falling observers}

The geodesic equations  (see Eq.  \eqref{eq:geodesic_equation}),  in the R\&W coordinate system \eqref{eq:rw},  take the form

\begin{equation}
\frac{d^2 \tau}{d \lambda^2}  + a^2 \mathcal{H} \left [  \left ( \frac{d y^{1}}{d \lambda} \right )^{\!2} +   \left ( \frac{d y^{2}}{d \lambda} \right )^{\!2}  + \left ( \frac{d y^{3}}{d \lambda} \right )^{\!2}   \right ] =0\,,\quad \; \frac{\d^2 y^{i}}{d \lambda^2} + 2 \mathcal{H} \tau' \frac{d y^{i}}{\d \lambda} =0\,, \label{eq:12}
\end{equation}
where $\mathcal{H}=1/a(da/d\tau)$ is the Hubble parameter, $i=1,2,3$ and  $\lambda$ is an  affine parameter.  The general  solutions of this system are given  in Eqs. \eqref{sys1}-\eqref{sys3}. The  equations of motion in the tilted time coordinate system \eqref{eq:line_element_gtz=q} are instead given  by
\begin{align}
& \frac{d^2 t}{d \lambda^2}  -q\frac{d^2z}{d\lambda^2}+ a^2 H \left [  \left ( \frac{d y^{1}}{d \lambda} \right )^{\!2} +   \left ( \frac{d y^{2}}{d \lambda} \right )^{\!2}  + \left ( \frac{d y^{3}}{d \lambda} \right )^{\!2}   \right ] =0\,,\quad \;   \frac{\d^2 x}{d \lambda^2} + 2 H \tau' \frac{d x}{\d \lambda} =0\,, \label{eq:12q}\\
& \frac{\d^2 y}{d \lambda^2} + 2  H \tau' \frac{d y}{\d \lambda} =0 \,, \quad \;
\frac{\d^2 z}{d \lambda^2} +\frac{q}{a^2}\frac{d^2 t}{d \lambda^2} + 2 H \tau' \frac{d z}{\d \lambda} =0\,, \label{eq:34q}
\end{align}
where $H=1/a (da/dt)$ is the Hubble parameter.  Incidentally, one may note that, as expected, the  comoving observers $x^i=const$ are in geodesic motion.  



\end{document}